# New Insights into Global Warming: End-to-End Visual Analysis and Prediction of Temperature Variations

**Meihua Zhou[1], Tianlong Zheng[1], Hanwen Xu[2], Li Yang[3], Tingting Wang[3,4]* and Nan Wan[3]***

---

**1. University of Chinese Academy of Sciences, China**

**2. School of Computer Science, Northeast Electric Power University, China**

**3. School of Medical Information, Wannan Medical College, China**

**4.School of Physics and Electronic Information, Anhui Normal University, China**

## Contributions

The study was conceptualized by M.Z., who also performed data visualization, time series modeling, clustering analysis, attribution analysis, and surveys. T.Z. assisted with code development, clustering analysis, and formatted the manuscript. H.X., N.W.and L.Y. supervised the manuscript proofreading. Supervision of the overall study was handled by N.W. and T.W. All authors critically revised the manuscript and endorsed the final version for publication.

## Corresponding author

Correspondence to: Tingting Wang[1] & Nan Wan[2]

ORCID ID[1]: 0000-0001-7565-1201 Email[1]: wangtt@wnmc.edu.cn

ORCID ID[2]: 0009-0008-8547-848X Email[2]: wannan@wnmc.edu.cn

## Abstract

Global warming presents an unprecedented challenge to our planet; however, comprehensive understanding remains hindered by geographical biases, temporal limitations, and lack of standardization in existing research. An end-to-end visual analysis of global warming using three distinct temperature datasets is presented. A baseline adjusted from the Paris Agreement's 1.5 °C benchmark based on data analysis is employed. A closed-loop design from visualization to prediction and clustering is created, using classic models tailored to data characteristics. This approach reduces complexity and eliminates the need for advanced feature engineering. A lightweight CNN-LSTM model specifically designed for global temperature change is proposed, achieving exceptional accuracy in long-term forecasting (MSE: $3\times10^{-6}$, $R^2$: 0.9999). DTW-KMeans clustering elucidates national-level temperature anomalies and carbon emission patterns. This comprehensive method reveals intricate spatiotemporal characteristics of global temperature variations and provides warming trend attribution. The findings offer new insights into climate change dynamics, demonstrating that simplicity and precision can coexist in environmental analysis.

## Introduction

Global warming represents one of the most pressing environmental challenges of our time[1, 2, 3]. The continuous rise in greenhouse gas emissions has led to a steady increase in Earth's average temperature,issues[4, 5, 6]. Extreme weather events, such as heatwaves and intense precipitation, pose significant threats to human health and ecosystem stability. The World Health Organization reports a year-on-year increase in heat-related mortality, with particularly pronounced effects among the elderly and children[7, 8]. As global warming intensifies, it is anticipated that the frequency and severity of extreme climate events will increase, thereby posing significant challenges to public health and global environmental systems [9, 10, 11].

While extensive research has focused on global warming and its impacts on climate and health[12, 13, 14], these studies often exhibit significant limitations. Firstly, there is a notable geographical bias, with many studies predominantly concentrating on specific regions such as North America or Europe [15, 16, 17, 18]. For example, V. Reyes-Garc´ıa et al.[19] highlight in their study that there is a paucity of geographic balance in the data from climate change studies, which is also reflected in the disparate directions of climate change studies. This narrow focus potentially overlooks crucial insights from other parts of the world, limiting our understanding of the global nature of climate change. Secondly, there is often a temporal limitation, with many studies focusing on short to medium-term impacts, neglecting long-term trend analysis crucial for comprehensive climate change understanding[20, 21]. The work of Miguel Molico[22] suffers from this drawback, as it focuses on the economic impacts of climate change only in the medium term.

Furthermore, the lack of a unified global standard for defining and categorizing temperature variations hampers the comparability and broad applicability of research findings[23, 24, 25, 26]. James et al.[27] reviewed methods for identifying regional climate responses to global warming targets and noted the methodological challenges in constraining regional climate signals due to varying data quality and standards across different studies Stein[28] reviewed the challenges at the intersection of climate science and statistics, pointing out that inconsistencies in data collection and standardization limit the ability to make accurate inferences about climate variability and change. This standardization issue is particularly problematic when attempting to synthesize results from various studies or apply findings across different geographical contexts. These limitations collectively impede a comprehensive understanding of global temperature changes and their multifaceted impacts on humanity and the environment.

To address these gaps, our study employs a multi-dimensional visual analysis approach using three distinct global temperature datasets. We examine global temperature trends, temperature anomalies, and other relevant metrics, utilizing the Paris Agreement's 1.5 °C benchmark as a reference point. Our analysis spans from 1880 to the present, with a baseline adjustment to account for pre-industrial warming, thus providing a more comprehensive temporal scope than many existing studies.

A key innovation in our approach is the integration of carefully selected time series models, including ARIMA(Autoregressive Integrated Moving Average)[29], SARIMA(Seasonal Autoregressive Integrated Moving Average)[30], and LSTM[31], alongside a novel CNN-LSTM(Convolutional Neural Network-Long

Short-Term Memory) model for predicting global temperature anomalies over the next century. Importantly, our model selection process prioritizes classic, lightweight structures that are well-suited to the specific characteristics of our data. This approach reduces model complexity and eliminates the need for advanced feature engineering, addressing a common limitation in existing studies that often rely on overly complex models requiring extensive data manipulation.

Our CNN-LSTM model, despite its simplicity, demonstrates exceptional performance, achieving remarkably low mean squared error[32] (MSE: $3\times10^{-6}$), mean absolute error[33] (MAE: 0.002), and a high R-squared value[34] ($R^2$: 0.9999). Similarly, for our clustering analysis, we employ the classic DTW-KMeans(Dynamic Time Warping - KMeans)[35] model, chosen for its effectiveness in handling time series data without requiring sophisticated feature engineering.

To the best of our knowledge, this study represents the first end-to-end visual analysis of global temperature changes. Our contributions include:

- A comprehensive visualization of global temperature trends, revealing spatiotemporal patterns in climate change across a broader geographical range than typically studied.
- The development and application of a novel CNN-LSTM model for accurate long-term temperature predictions, demonstrating that lightweight, classic models can yield high-performance results when properly aligned with data characteristics.
- An in-depth attribution analysis of global warming causes based on our predictive models, offering insights into long-term trends often overlooked in shorter-term studies.
- A time series clustering analysis of temperature anomalies and carbon emissions at the national level using DTW-KMeans, providing nuanced insights for policy-making and addressing the lack of standardization in temperature variation categorization.

## Visualization of data features

---

### Patiotemporal Characteristics of Data

Fig.1 and Supplementary Figure 1 provide complementary perspectives on the global distribution of temperature anomalies, collectively elucidating the spatiotemporal dynamics of climate change and offering a vivid glimpse into the repercussions of global warming. Fig.1, depicting national temperature variations with color-coded anomalies, reveals a pronounced warming trend in North America, Europe, and parts of Asia. Supplementary Figure 1 enhances spatial acuity with a three-dimensional spherical projection, highlighting the geographic extent of these changes. The Northern Hemisphere, particularly the Arctic, experiences the most dramatic shifts. This detailed perspective offers a nuanced view of anomalies in Europe, Africa, and the Middle East, especially in central and eastern Europe, northern and central Africa, and the desert and coastal regions of the Middle East. These anomalously warm regions are linked to an increase in extreme weather events, portending heightened climate risks for these areas in the future.

**Fig. 1: Global temperature change (flat map view).**

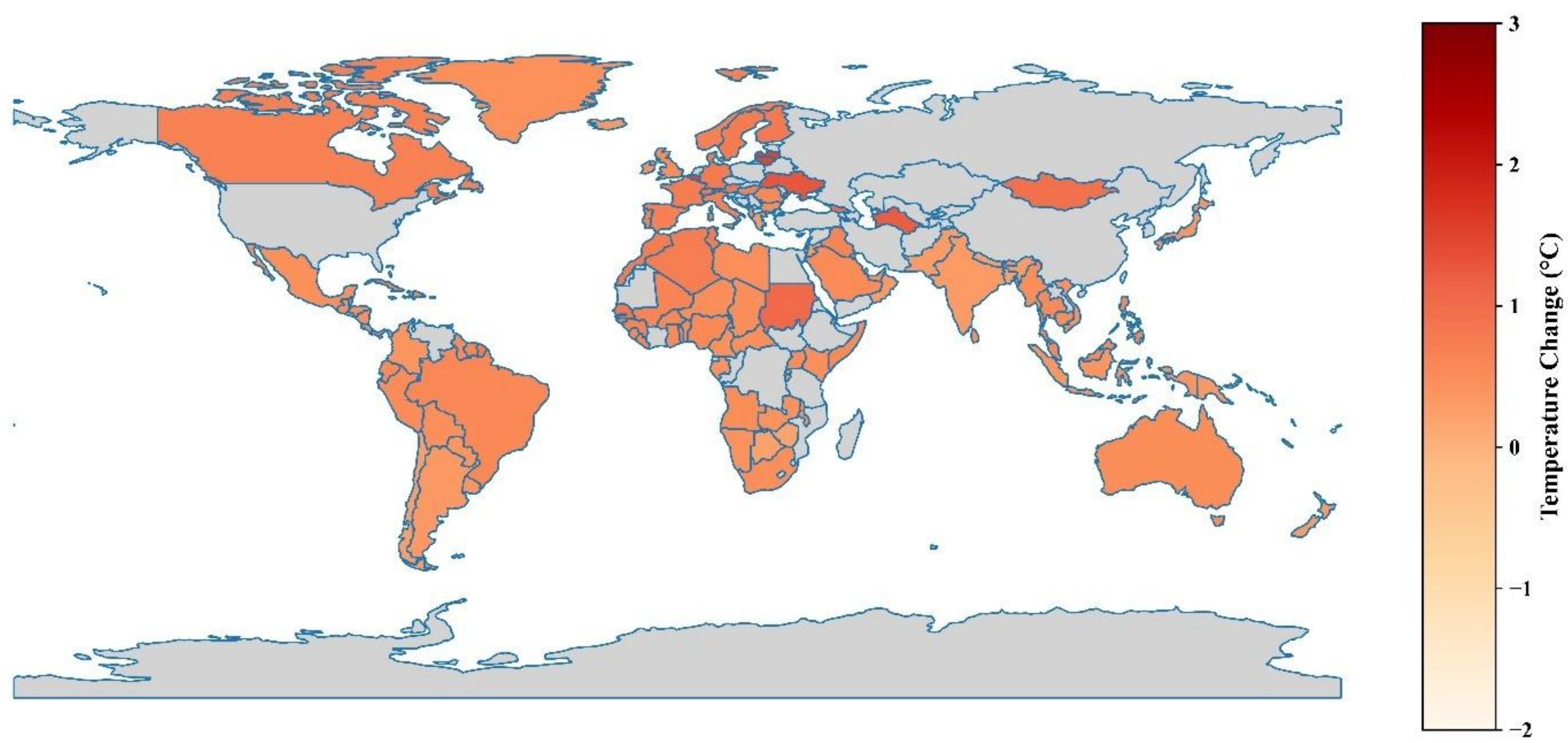

The map shows temperature variations across countries, with colors representing changes in degrees Celsius, ranging from -2°C to +3°C. Red indicates increases, while grey denotes missing data.

Supplementary Figure 2 depicts the temporal trend of global populations affected, killed, and injured by extreme climate events, including heatwaves, from 1999 to 2023. This visualization illustrates the profound impact of such events on human health, with significant peaks in affected and deceased individuals occurring in 2000, 2003, 2010, and 2015. The data demonstrate that global warming presents not only a thermal challenge but also a significant threat to human well-being. Understanding these shifts is crucial for developing effective public health and emergency response strategies.

The collective data presented in these figures demonstrate the disparate effects of global warming across regions. They highlight the unequal distribution of temperature anomalies and the resulting challenges for formulating global climate policy. Therefore, tailored strategies informed by regional specifics are imperative to address these varied impacts effectively.

## Characteristics of Data for Prediction

Supplementary Figure 3 elucidates the characteristics and trends of the global temperature dataset for prediction purposes. Supplementary Figure 3a presents a time series of temperature anomalies, with colors coding for monthly and seasonal variations. It reveals distinct temporal patterns, particularly pronounced in summer and winter, reflecting the seasonality of global climate. Supplementary Figure 3b, a Q-Q plot, assesses the normality of the temperature data, showing that the data follows a largely normal distribution in the central range, with deviations at the extremes. These deviations indicate nonlinear characteristics and the presence of outliers, presenting challenges for prediction models while offering crucial statistical insights into the data's structure.

Supplementary Figure 4 displays temperature distributions, median changes, and extreme ranges for each month and season, as indicated by box plots. It highlights the variability and central tendencies of temperature anomalies, revealing the periodic and local features of temperature changes.

Fig.2 illustrates long-term trends in temperature anomalies. Fig.2a depicts seasonal variations in temperature, while Fig.2b presents a ten-year moving average and linear trend line, revealing a century-long warming trend, particularly accelerated in recent decades. These trends are vital for forecasting future temperature changes.

**Fig. 2: Historical temperature anomalies from 1880 to 2020.**

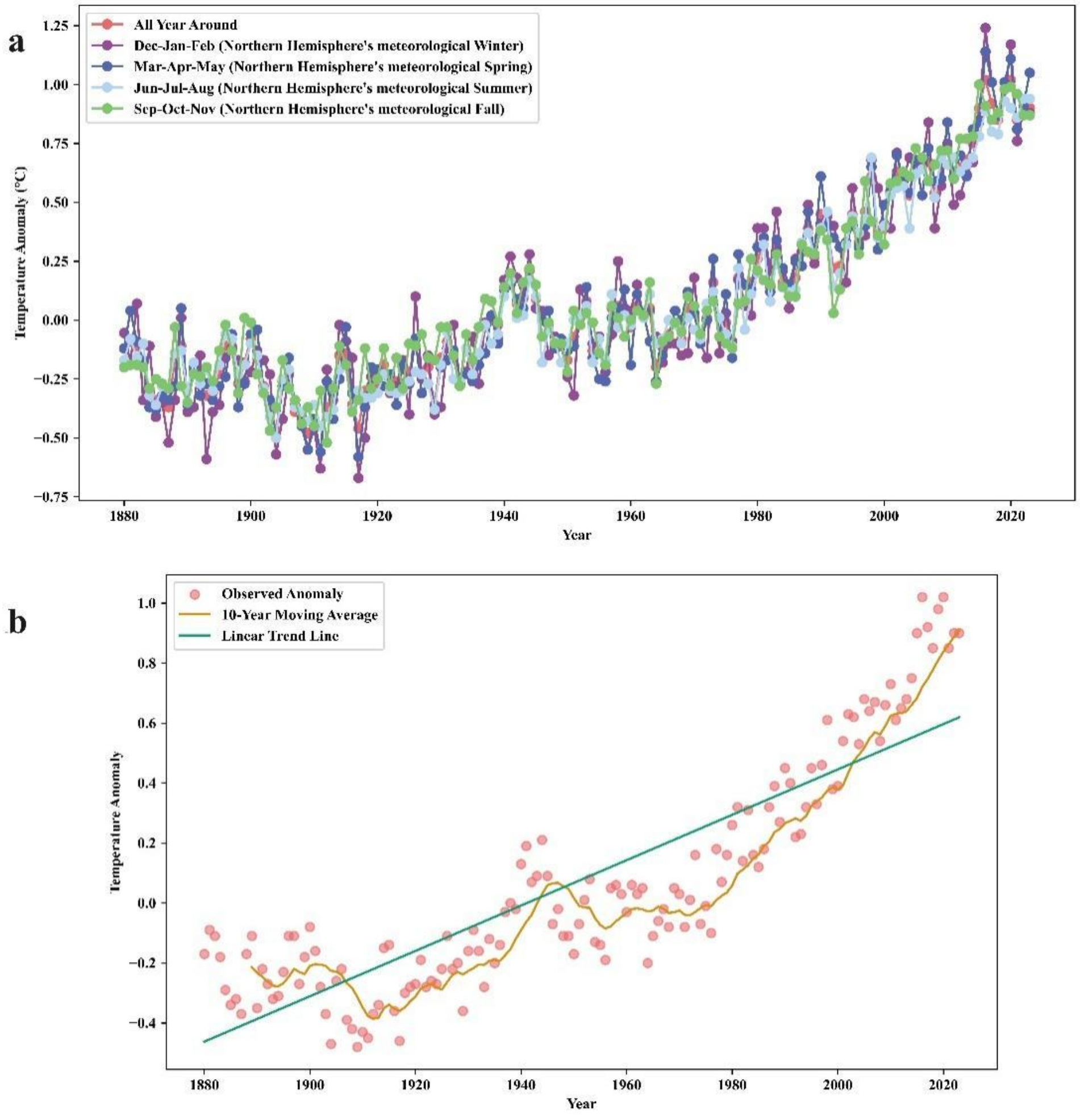

**a** Seasonal temperature anomalies: all year round, winter, spring, summer, and fall. **b** Temperature anomalies, 10-year moving average, and linear trend line.

In a nutshell, the dataset exhibits distinct temporal characteristics, including monthly and seasonal variations, and periodic features that reflect both long-term and short-term temperature trends. Its relatively straightforward structure supports the use of classic, lightweight prediction models, minimizing the need for complex feature engineering. Additionally, it captures long-term trend changes and notable local features, encompassing short-term fluctuations and anomalies.

Given the temporal and spatial characteristics of the dataset, along with the long-term and short-term periodic variations, prediction models must capture these features. Such models must handle both linear and nonlinear features, as well as cyclical and seasonal changes, to provide accurate predictions.

# Results

## Prediction Model Performance

The ability to forecast global temperatures is crucial for understanding climate change and informing policy decisions. This study compares various time series models, including a novel convolutional neural network (CNN) architecture combined with a long short-term memory (LSTM) unit, which demonstrated superior performance in predicting temperature trends. Detailed model performance is presented in Table1.

The global temperature dataset exhibits complex characteristics, such as pronounced temporal patterns, periodic features reflecting both long-term and short-term trends, and significant local fluctuations. These attributes require a model capable of capturing multi-scale temporal dynamics while maintaining computational efficiency.

The CNN-LSTM model addresses these challenges through a meticulously designed architectural framework, depicted in Fig.4a. The convolutional layer, comprising 32 filters with a kernel size of 3, effectively captures local temporal patterns and short-term fluctuations. This is followed by a ReLU activation function, introducing non-linearity to model complex data relationships. The subsequent LSTM layer, with 50 hidden units, processes the CNN-extracted features, maintaining sensitivity to long-term dependencies and trends. Finally, a linear output layer maps the LSTM output to temperature predictions.

This architecture's strength lies in its synergistic combination of CNN and LSTM components. The CNN efficiently extracts local temporal features without pooling, preserving fine-grained information. The LSTM then processes these features, maintaining temporal order sensitivity crucial for long-term trend analysis. This approach balances model complexity and predictive power, aligning well with the data's structure and eliminating the need for complex feature engineering.

**Fig. 3: Validation on historical data.**

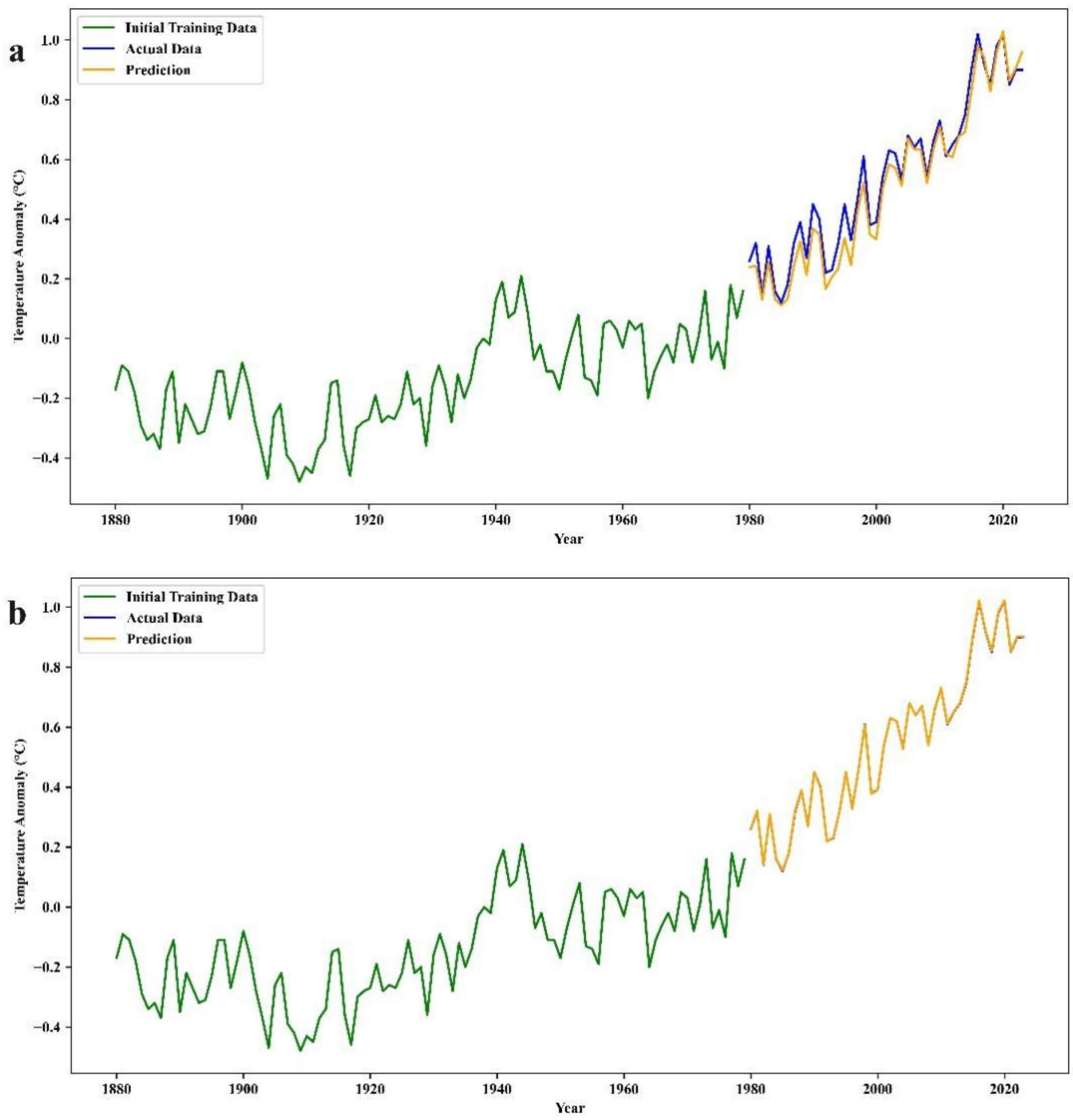

**a** Temperature anomaly predictions using LSTM. **b** Temperature anomaly predictions using CNN-LSTM.

Validation on historical data (Fig.3) demonstrates the CNN-LSTM model's superior ability to capture both long-term trends and short-term variations compared to traditional LSTM models. This improved accuracy is critical for reliable climate projections and policy guidance. Using this optimized model, we projected global temperature trends for the next century (Fig.4b). After adjusting our baseline to align with the Paris Agreement's 1.5 °C target, our model predicts that global temperatures may reach the critical +1.0°C threshold (equivalent to the Paris Agreement's 1.5 °C target) by 2031. This projection underscores the urgency of implementing effective climate mitigation strategies.

**Fig. 4: CNN-LSTM model architecture and temperature prediction.**

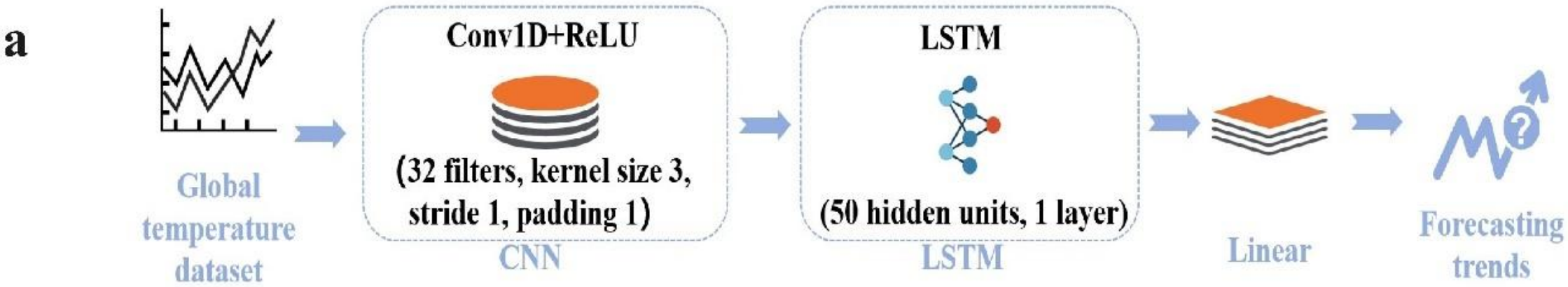

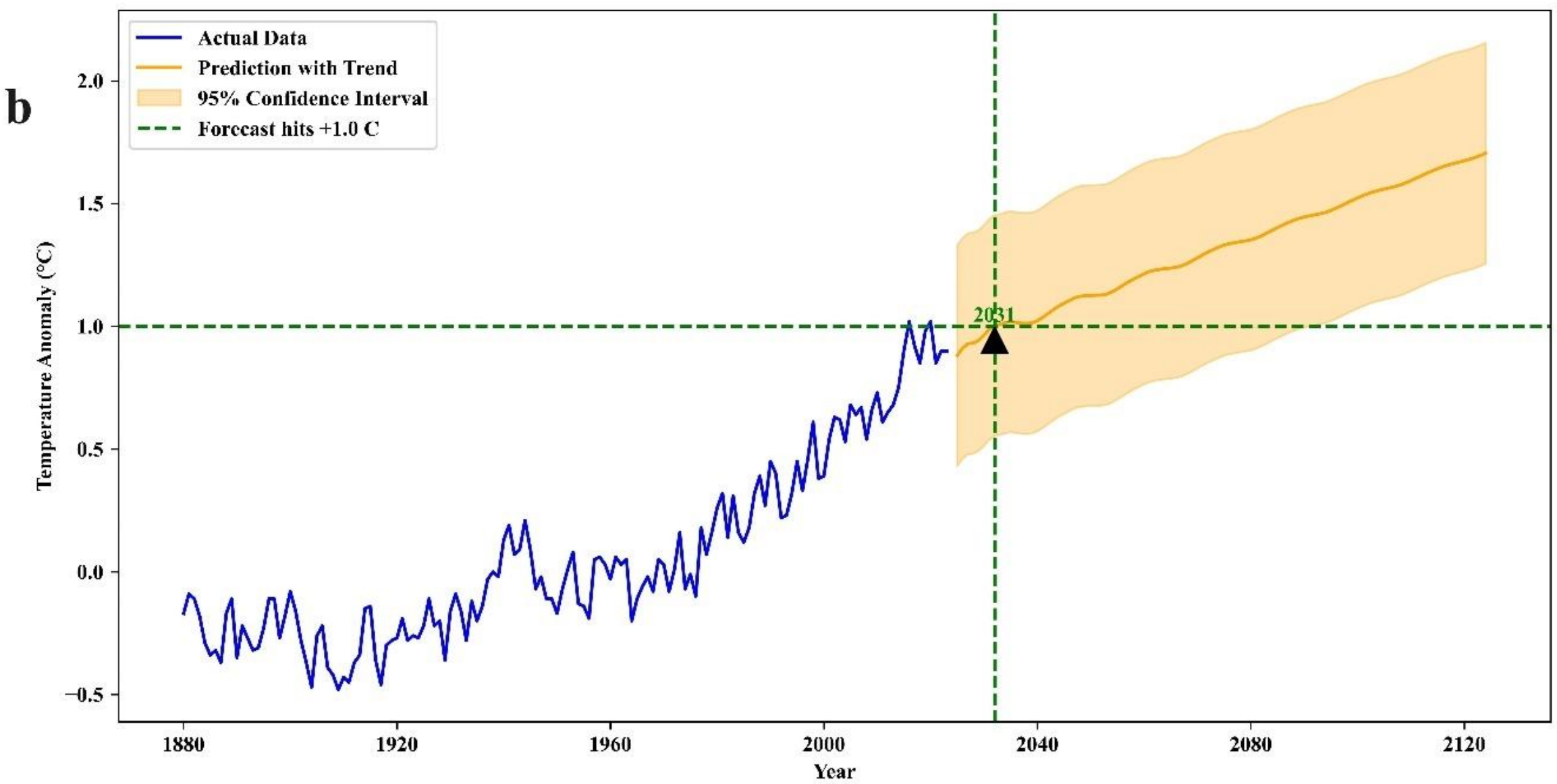

**a** Proposed CNN-LSTM model architecture. **b** Predicted temperature changes for the next 100 years, with a baseline adjusted to 1.0°C according to the Paris Agreement.

In conclusion, the superior performance of our CNN-LSTM model in global temperature prediction stems from its ability to effectively capture multi-scale temporal features without sacrificing computational efficiency. This approach not only provides accurate short-term predictions but also offers reliable long-term forecasts, crucial for informing climate policy and action. The model's projection of imminent critical temperature threshold crossing emphasizes the need for immediate and sustained efforts to mitigate global warming.

## Attribution Analysis of Global Warming

### Impact of Energy Use and Emissions

The primary drivers of global warming can be elucidated through a comprehensive analysis of energy consumption patterns, greenhouse gas emissions, and natural climate variability. Fig.5 illustrates $CO_2$ emissions from various energy sources from 2000 to 2023, showing that fossil fuel utilization remains the predominant source of $CO_2$ emissions, exhibiting a persistent upward trajectory. This trend correlates directly with the increase in radiative forcing from greenhouse gases shown in Fig.6, where $CO_2$ makes the most significant contribution.

**Fig. 5: $CO_2$ emissions from electricity generation by source from 2000 to 2023.**

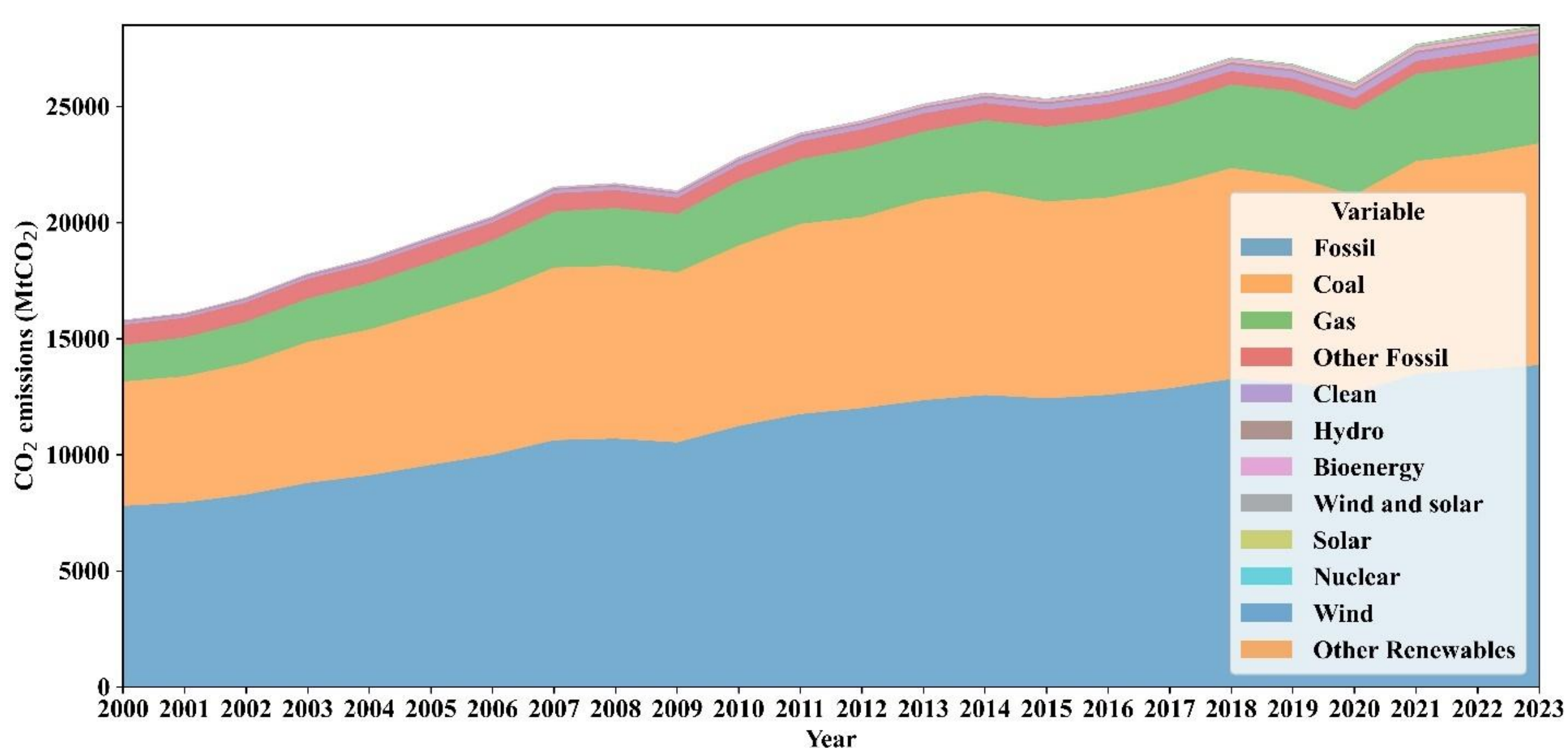

Emissions from fossil fuels, coal, gas, other fossil sources, clean energy, hydro, bioenergy, wind and solar, solar, nuclear, wind, and other renewables.

The evolving patterns of energy utilization not only alter atmospheric composition but also influence the global climate system through complex feedback mechanisms. Fig.6a, depicting temperature anomalies associated with El Niño and La Niño events, exemplifies the short-term impacts of natural climate variability on global temperatures. However, these natural fluctuations are superimposed upon the long-term warming trend driven by anthropogenic greenhouse gas emissions, underscoring the dominant role of human activities in shaping the climate system.

**Fig. 6: Oceanic Niño Index and radiative forcing of greenhouse gases.**

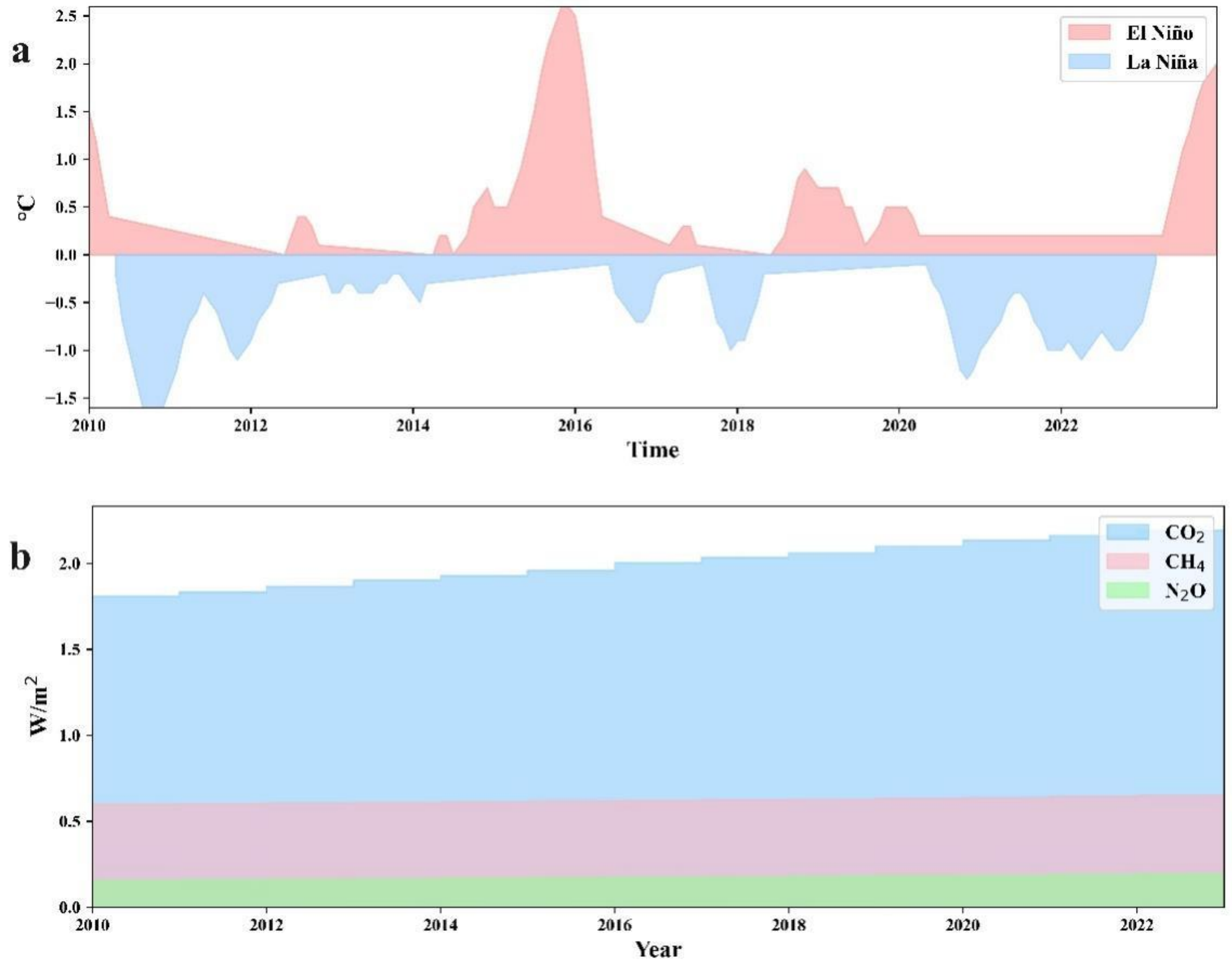

**a** Oceanic Niño Index from 2010 to 2023. **b** Radiative forcing of greenhouse gases $CO_2$, $CH_4$, and $N_2O$.

Notably, the radiative forcing of $CO_2$, $CH_4$, and $N_2O$, as shown in Fig.6b, all exhibit upward trends, aligning closely with the increased fossil fuel usage depicted in Fig.5. This congruence reinforces the causal relationship between energy utilization and global warming while simultaneously highlighting potential pathways for climate change mitigation through a transition to low-carbon energy systems.

Deforestation and Climate Implications

Deforestation emerges as another critical driver of global warming, with complex and far-reaching impacts. Fig.7a reveals the primary causes of deforestation, with commodity-driven deforestation (20.9%) and shifting agriculture (23.2%) being the predominant factors. This not only directly diminishes the Earth's carbon sequestration capacity but also indirectly affects the climate system by altering surface albedo and evapotranspiration processes.

**Fig. 7: Global environmental changes.**

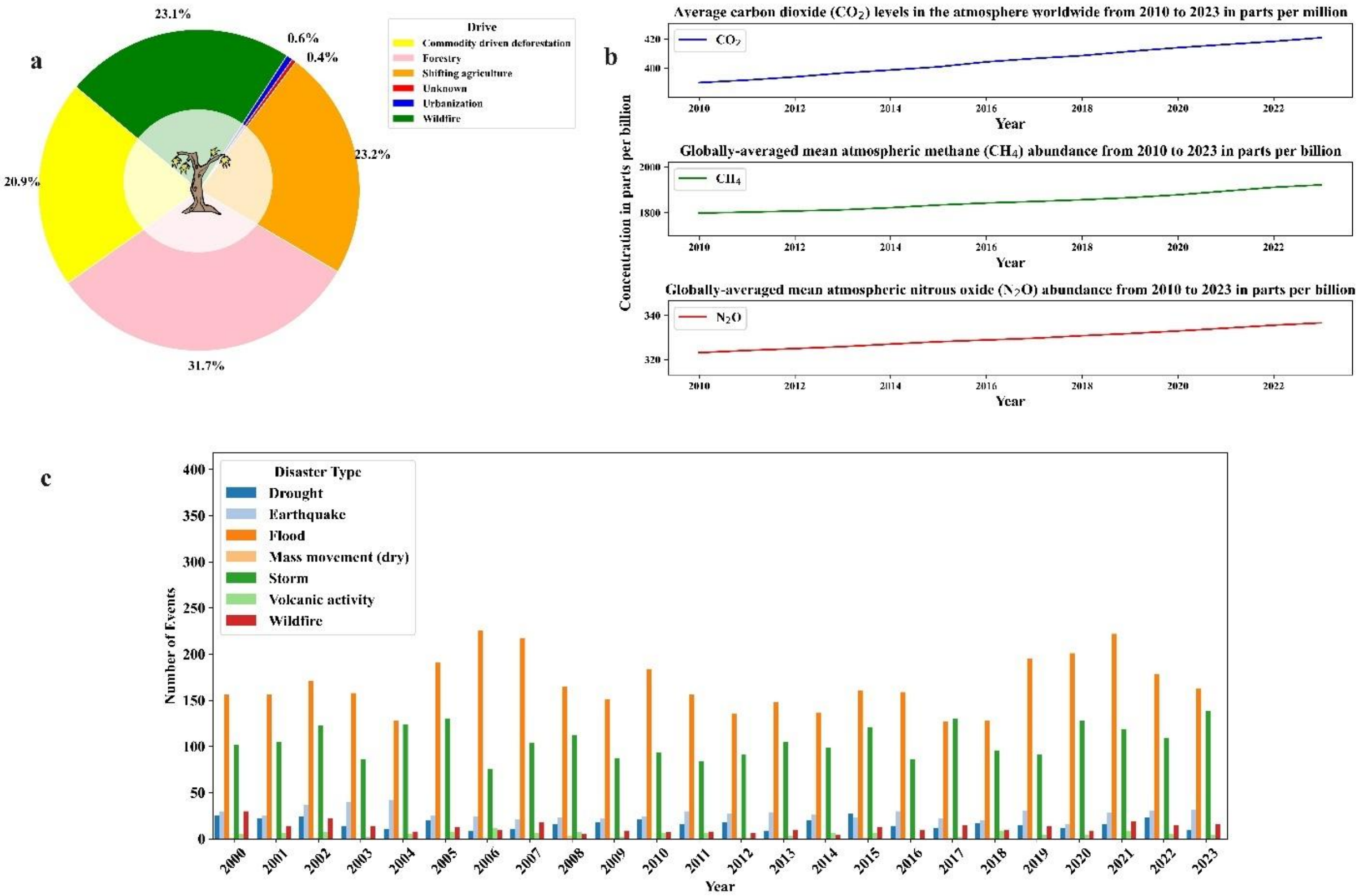

**a** Global tree cover loss from 2002 to 2023. **b** Globally-averaged mean atmospheric $CO_2$, $CH_4$, and $N_2O$ abundance from 2010 to 2023 in parts per billion. **c** Global extreme weather events from 2000 to 2023.

The climate implications of deforestation are corroborated by the changes in greenhouse gas concentrations illustrated in Fig.7b. Between 2010 and 2023, concentrations of $CO_2$, $CH_4$, and $N_2O$ all show ascending trends, closely linked to carbon emissions from deforestation and land-use changes. The persistent rise in $CO_2$ concentration, in particular, reflects the weakening of forests as carbon sinks.

Fig.7c demonstrates an increase in the frequency of extreme weather events, especially floods and storms, which can be interpreted as a direct consequence of global warming. This trend correlates strongly with the rising greenhouse gas concentrations shown in Fig.7b, indicating that deforestation indirectly exacerbates the occurrence of extreme weather events by increasing greenhouse gas emissions and altering local climate conditions.

Undoubtedly, changes in energy utilization patterns and deforestation are the primary drivers of global warming. This highlights the intricate interplay between human activities and climate change, underscoring the urgent need for integrated approaches to address the multifaceted challenges posed by global warming. The continued use of fossil fuels increases atmospheric concentrations of greenhouse gases, while deforestation exacerbates this process by reducing carbon sinks and altering surface characteristics. The combined effect of these factors not only accelerates global warming but also increases the frequency and intensity of extreme weather events. Therefore, mitigating global warming necessitates a dual strategy: transitioning to sustainable energy systems and enhancing forest protection and restoration to maintain Earth's ecological balance and climate stability.

## Time Series Clustering

Building upon the comprehensive attribution analysis of global warming drivers, this study employs advanced time series clustering techniques to elucidate the spatiotemporal dynamics of temperature anomalies and carbon emissions. By leveraging Dynamic Time Warping (DTW) coupled with K-Means clustering to analyze data from 1860 to 2023, we aim to uncover intricate relationships between these climate indicators and geographical factors, thereby providing a more nuanced understanding of how anthropogenic influences manifest across different regions globally.

The global climate system's response to anthropogenic forcing exhibits complex patterns that vary significantly across space and time.Traditional attribution studies, while crucial for identifying overarching drivers of climate change, often lack the granularity needed to capture regional variations and the temporal evolution of climate indicators. To address this limitation, our clustering analysis seeks to reveal distinct regional patterns in both temperature anomalies and carbon emissions, offering insights into how the primary drivers of global warming identified in attribution studies translate into observable climate trends across diverse geographical contexts.

This approach allows us to explore how energy utilization patterns and land-use changes, particularly deforestation, manifest in terms of carbon emissions and temperature anomalies across different parts of the globe. By doing so, we can better understand the differential impacts of these drivers on various regions, accounting for factors such as the level of industrialization, economic development trajectories, and geographical characteristics. This nuanced perspective is essential for developing targeted mitigation strategies and adapting global climate policies to regional contexts.

Temperature anomaly clustering yielded three distinct patterns (Fig.8a), each corresponding to specific latitudinal bands. Cluster 0, predominantly associated with high-latitude regions, demonstrates a relatively stable temperature profile with moderate warming. Cluster 1, characteristic of mid-latitude countries, exhibits more pronounced temperature increases, particularly post-1970. Cluster 2, primarily representing low-latitude areas, shows the most dramatic temperature rise, especially since the 1980s. This latitudinal gradient in temperature anomalies (Supplementary Figure 5a) aligns with established climate models predicting polar amplification and differential warming rates across latitudes.

**Fig. 8: Cluster analysis of temperature anomalies and $CO_2$ emissions.**

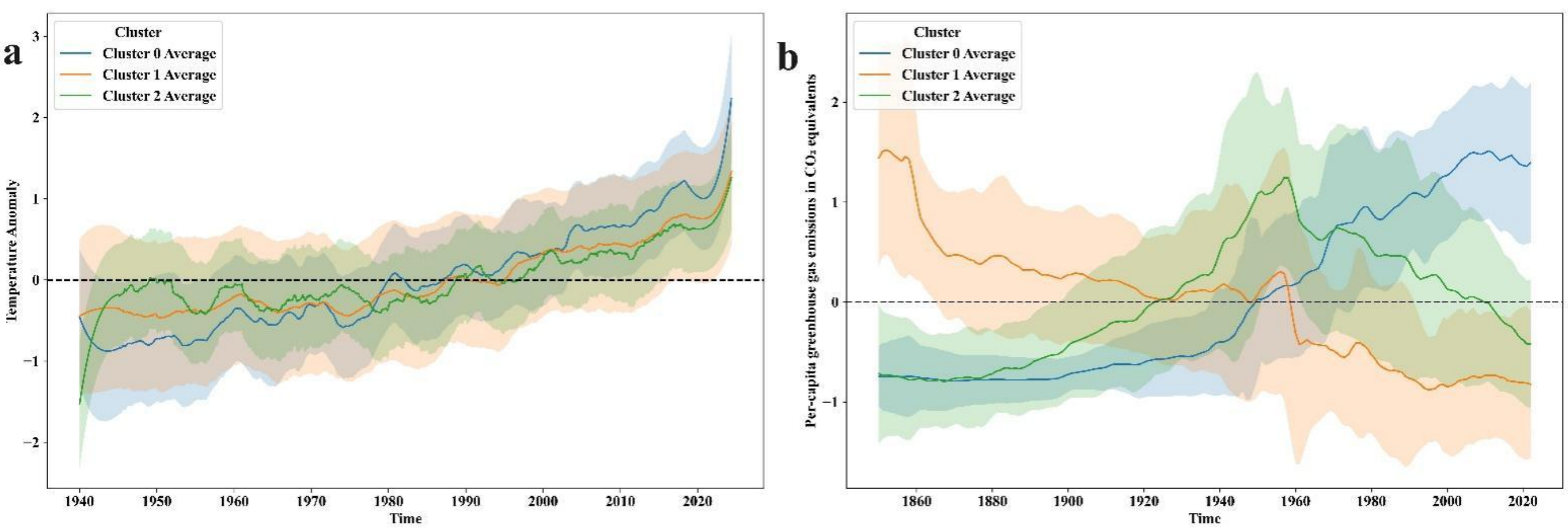

**a** Temperature anomalies for different clusters of countries from 1940 to 2020. **b** Per-capita greenhouse gas emissions in $CO_2$ equivalents for different clusters of countries from 1800 to 2020.

Carbon emission clustering (Fig.8b) reveals more pronounced temporal disparities, reflecting the heterogeneous nature of global industrialization and economic development. Cluster 0, typical of developed nations, exhibits early industrialization peaks followed by stabilization or decline, indicative of advanced environmental policies and technological progress. Cluster 1, characteristic of rapidly industrializing nations, shows a sharp increase in emissions post-1950, coinciding with accelerated economic growth. Cluster 2, representative of emerging economies, demonstrates a delayed but steep rise in emissions, particularly in recent decades. The geographical distribution of these emission clusters (Supplementary Figure 5b) underscores the complex interplay between economic development stages, technological advancement, and environmental policies across different regions.

Juxtaposition of temperature anomaly and carbon emission clusters reveals compelling correlations. High-emission regions (Cluster 2 in emissions) often correspond to areas with significant temperature anomalies (Cluster 2 in temperature), particularly in low-latitude developing nations. This relationship suggests a direct link between local emissions and regional temperature increases, although global atmospheric circulation complicates this association. Moderate-emission areas (Cluster 1 in emissions) generally align with intermediate temperature anomalies (Cluster 1 in temperature), typical of mid-latitude countries. Low-emission zones (Cluster 0 in emissions) frequently coincide with regions of lower temperature anomalies (Cluster 0 in temperature), characteristic of high-latitude developed nations, although this relationship is confounded by polar amplification effects.

**Fig. 9: Correlation matrix and temperature contour map.**

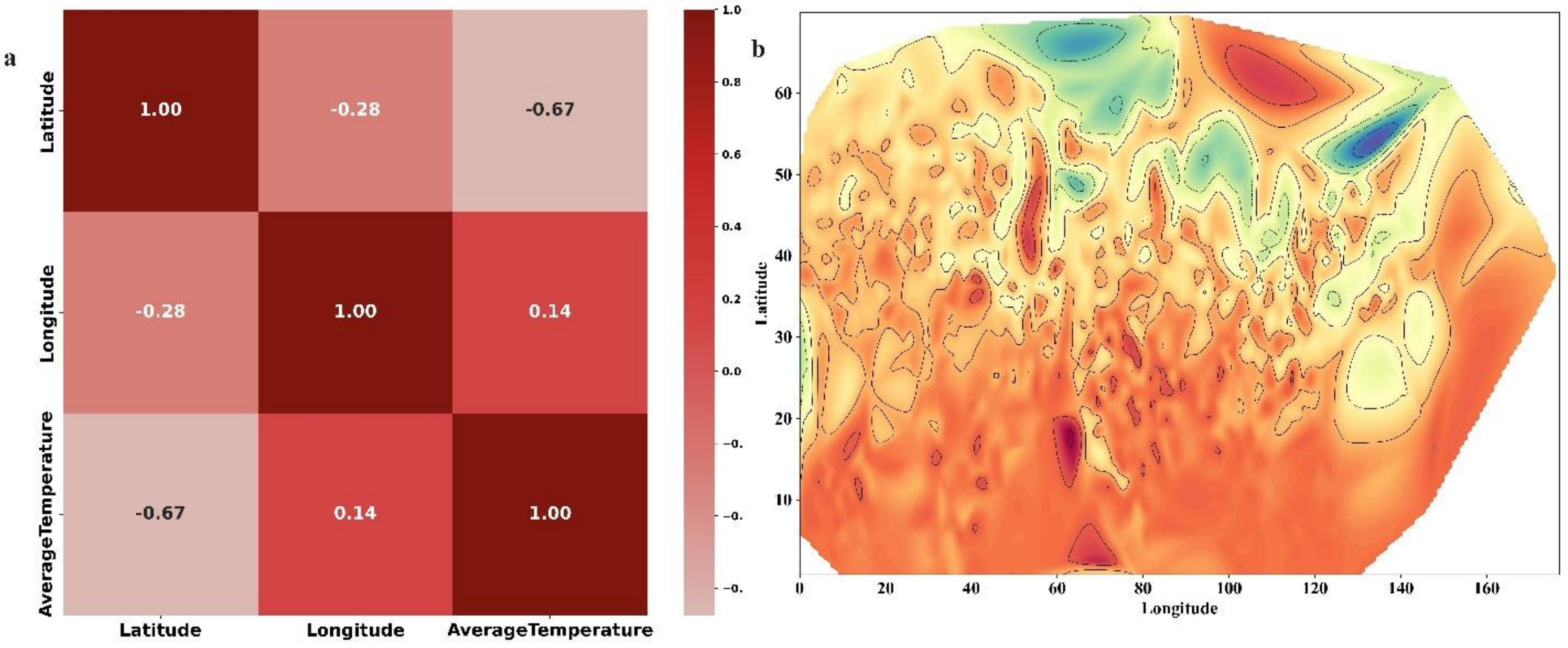

**a** Correlation matrix of latitude, longitude, and average temperature. **b** Contour map of average temperature distribution.

The confusion matrix (Fig.9a) reveals a strong negative correlation (-0.67) between latitude and average temperature, whereas the longitudinal correlation is notably weaker (0.14). This geographic asymmetry in temperature patterns is visually reinforced by the temperature contour map (Fig.9b), which clearly delineates latitudinal temperature gradients. These findings underscore the critical role of geographical factors in modulating the relationship between carbon emissions and temperature anomalies. This analysis sheds light on the complex, non-linear relationships among carbon emissions, temperature anomalies, and geographical factors. The observed patterns highlight the urgent need for tailored mitigation strategies that account for regional disparities.

## Discussion

This study introduces a groundbreaking approach to global warming research, offering novel perspectives and insights through an innovative end-to-end visualization and analysis methodology. By integrating machine learning techniques with established climate science methodologies, we have developed a comprehensive analytical framework that encompasses diverse aspects of temperature data visualization, attribution analysis, regional clustering, and future projections. This integrated approach not only mitigates the geographical bias and time constraints inherent in existing studies but also provides a holistic framework for understanding the complex dynamics of global warming.

Our visualizations reveal extensive spatiotemporal patterns of climate change, offering an intuitive grasp of global warming and facilitating access to complex climate data. The hybrid CNN-LSTM model for temperature forecasting challenges the assumption that only complex models are essential for climate prediction, demonstrating that simple, well-designed models can effectively capture both short- and long-term climate trends. Projections for the next century indicate an acceleration of global warming, with average temperatures potentially surpassing pre-industrial levels by 1.5°C as early as 2031 under high-emission scenarios, underscoring the urgency of tailored climate strategies.

Attribution analysis identifies energy use and deforestation as key human drivers of warming, while DTW-KMeans clustering reveals regional climate patterns, emphasizing the importance of latitude in climate response and informing targeted mitigation strategies. The innovation of this study lies not only in its methodology but also in its significant contributions to climate science and policy-making. By integrating global attribution and regional clustering analyses, we bridge the gap between macro-level driving factors and micro-level climate change manifestations.

All in all, this paper is the first to propose an end-to-end visualization analysis, establishing a closed-loop analytical approach. This comprehensive method lays the foundation for future research and more targeted climate policies, promising to develop more effective strategies for mitigating and adapting to climate change at both global and regional scales. Looking ahead, we plan to incorporate additional climate variables and develop higher-resolution regional climate models to deepen our understanding of the complex dynamics of global warming. The application of this model demonstrates its potential to provide critical insights into climate change, aiding policymakers and scientists in formulating more nuanced and effective strategies.

# Methods

## Datasets

This paper employs a multitude of comprehensive databases to facilitate the end-to-end visualization and analysis of global temperature changes. The initial dataset employed is the “Climate Change Indicators” dataset from Kaggle (1961-2022). This offers detailed climate change indicators and country codes, with annual updates ensuring the information’s timeliness. The dataset can be accessed via the following link: https://www.kaggle.com/datasets/tarunrm. Subsequently, the “Earth Surface Temperature Data” from the Berkeley Earth Surface Temperature Study is employed, incorporating 1.6 billion temperature records and integrating additional short-term meteorological observations through specialized methods (Dataset link: https://berkeleyearth.org/data/). The aforementioned datasets facilitate a comprehensive analysis of temperature data characteristics. To analyze the causes of global warming, a Kaggle dataset is employed, which comprehensively examines temperature, carbon dioxide emissions, and other key climate indicators to identify the primary drivers of global warming (Dataset link: https://www.kaggle.com/datasets/uyenlex/vda-global-warming). In cluster analysis, the global temperature anomaly dataset based on ECMWF ERA5 project data (using 1991-2020 as the baseline) is employed, along with the carbon dioxide and greenhouse gas emissions dataset for each country from Our World in Data. This diverse collection of datasets provides a robust foundation for our multi-dimensional analysis (Dataset links: https://ourworldindata.org/temperature-anomaly and https://ourworldindata.org/$CO_2$-and-greenhouse-gas-emissions).

## Baseline Adjustment for Prediction

The Paris Agreement, ratified in 2015, aims to limit global warming to within 1.5 degrees Celsius above pre industrial levels[36, 37, 38, 39]. This science-based threshold indicates that maintaining global temperatures below this level will significantly mitigate the risks and impacts of climate change. The World Resources Institute has conducted a crucial analysis of the dangers associated with the 1.5 °C and 2 °C thresholds mentioned in the Paris Agreement, as illustrated in Supplementary Figure 6. This analysis elucidates the far-reaching effects of exceeding these temperature limits and underscores the importance of the Paris Agreement's goals. Such comprehensive studies on the potential impacts at different warming levels underpin the use of the Paris Agreement as the primary reference standard in this paper.

The Paris Agreement does not specify a particular pre-industrial period for calculating temperature anomalies, though the years 1850 to 1900 are widely accepted by the scientific community as a reliable baseline. The dataset used in this paper for predictions starts in 1880 and uses the period 1951-1980 as the reference for temperature anomalies. Since this reference period is not pre-industrial, it is necessary to adjust the temperature anomaly baseline to accurately reflect the difference from pre-industrial levels. The period 1951-1980 is already warmer than the pre-industrial baseline, so adding 1.5 °C directly does not correspond to the Paris Agreement's target. Visual inspections and estimates presented in this paper indicate that the global mean temperature in the mid-20th century was approximately 0.3°Cto 0.4 °C higher than the pre-industrial level. Considering this discrepancy, the baseline is adjusted downwards by 0.5 °C , with +1.0 °C used as the revised baseline in this analysis.

## Model Selection

In this paper, a thorough statistical visualization analysis of the datasets revealed a relatively simple data structure, prompting the selection of classic and commonly used models for both prediction and clustering analyses. This approach minimized model complexity and obviated the need for sophisticated feature engineering.

For time series prediction, we evaluated the ARIMA, SARIMA, and LSTM models. ARIMA is a statistical model used for time series forecasting. It combines Autoregressive (AR), Integrated (I), and Moving Average (MA) components to capture the temporal dependencies in the data. The basic equation is as follows:

$$\begin{aligned} Y_t = \phi_1 Y_{t-1} + \phi_2 Y_{t-2} + \ldots + \phi_p Y_{t-p} \\ + \theta_1 \epsilon_{t-1} + \theta_2 \epsilon_{t-2} + \ldots + \theta_p \epsilon_{t-p} + \epsilon_t \end{aligned} \quad (1)$$

where $Y_t$ is the current value of the time series, $\phi_i$ and $\theta_j$ are the autoregressive and moving average coefficients, respectively, and $\epsilon_t$ is the white noise.

SARIMA extends the ARIMA model by incorporating seasonal components to capture the seasonal fluctuations in the data. The equation is:

$$Y_t = \phi(B) y_{t-1} + \Theta(B^s) \epsilon_{t-s} + \epsilon_t \quad (2)$$

where $\phi(B)$ and $\Theta(B^s)$ are polynomials representing non-seasonal and seasonal components, respectively, and $B$ is the backshift operator.

Our results indicated that the LSTM model outperformed the others in capturing global temperature variability. The LSTM model's ability to handle long-term dependencies in sequence data, coupled with its overcoming of the vanishing gradient problem inherent in traditional RNNs (Recurrent Neural Networks)[40], makes it particularly suitable for long-term climate data prediction, despite some shortcomings in local feature handling. LSTM uses hidden states $h$ and cell states $c$ to regulate the flow of information. The equations at time step $t$ are:

$$
\begin{cases}
\tilde{c}_t = \tanh(W^c h_{t-1} + I^c x^t) \\
i_t = \sigma(W^i h_{t-1} + I^i x^t) \\
f_t = \sigma(W^f h_{t-1} + I^f x^t) \\
o_t = \sigma(W^o h_{t-1} + I^o x^t) \\
c_t = f_t \odot c_{t-1} + i_t \odot \tilde{c}_t \\
h_t = o_t \odot \tanh(c_t)
\end{cases} \qquad (3)
$$

where: $\tilde{c}_t$ is the candidate cell state, $i_t$ is the input gate, $f_t$ is the forget gate, $o_t$ is the output gate, $c_t$ is the cell state, $h_t$ is the hidden state, $W^c$, $W^i$, $W^f$, $W^o$ are weight matrices for the cell state, input gate, forget gate, and output gate, respectively, $I^c$, $I^i$, $I^f$, $I^o$ are input weight matrices.

However, the LSTM model introduces noise that is unrelated to temperature prediction and is also affected by larger and smaller values of short-term fluctuations in the time-series data. This results in poor predictions.

Inspired by the success of CNN-LSTM models in autonomous driving[41, 42, 43], where they excel in capturing the spatiotemporal dynamics of traffic participants, we adapted this approach for global temperature change prediction[44]. CNNs are adept at processing image data and can extract local patterns and features from time series through local convolution operations, which is crucial for capturing short-term temperature trends[45, 46, 47]. By leveraging CNNs[48] to extract local time series features and LSTM to model global temporal dependencies, we aimed to enhance prediction accuracy and robustness[49, 50].To align with our data characteristics, we designed a simple CNN-LSTM structure that eliminated the need for complex feature engineering. The proposed CNN-LSTM model's computational equations and pseudocode are provided below to illustrate the implementation.

For clustering analysis, we employed the DTW-KMeans algorithm, a hybrid of Dynamic Time Warping (DTW)[51] and KMeans[52], which effectively clusters time series data, especially when dealing with sequences of varying lengths and nonlinear transformations.The pseudocode for the DTW-KMeans algorithm is as follows:

| **Algorithm 1: CNN-LSTM Model for Climate Prediction** |
| --- |
| |

**Input:** Input sequence x of shape (batch size, sequence length, features)

Initialize parameters: $W_{conv}, b_{conv}, W_f, b_f, W_i, b_i, W_C, b_C, W_o, b_o, W_{fc}, b_{fc}$

Function CNN-LSTM(x)

//Convolutional layer

$z = Conv1d(x, W_{conv}) + b_{conv}$

$a = ReLU(z)$

// Reshape for LSTM

$a = Permute(a)$

//LSTM layer

for $t = 1$ to $T$ do

$f_t = sigmoid\left(W_f * [h_{t-1}, a_t] + b_f\right)$

$i_t = sigmoid(W_i * [h_{t-1}, a_t] + b_i)$

$\tilde{C}_t = tanh(W_C * [h_{t-1}, a_t] + b_C)$

$C_t = f_t \odot C_{t-1} + i_t \odot \tilde{C}_t$

$o_t = sigmoid(W_o * [h_{t-1}, a_t] + b_o)$

$h_t = o_t \odot \tanh(C_t)$

end for

// Fully connected layer

$y = W_{fc} * h_T + b_{fc}$

return y

end function

**Output:** Predicted climate value y

**Algorithm 2: DTW-KMeans for Time Series Clustering**

**Input:** Set of time series data $X$,number of clusters $K$

```
    Function DTW-KMeans(X, K)
    //Step l: Compute DTW distance matrix
    D=compute DTW distance matrix(X)
    //Step 2: Initialize KMeans
    centroids=initialize centroids(D, K)
    repeat
        // Step 3: Assign clusters based on DTW distance
         clusters= assign clusters(X, centroids, D)
        // Step 4: Update centroids based on cluster assignments
         centroids=update centroids(clusters, D)
     until centroids do not change
     return clusters
end Function
```

**Output:** Cluster assignments

## Code Availability

Our primary code and weight are available at https://github.com/mh-zhou/new.

## Ethics declarations

### Competing interests

The authors declare no competing interests.

## Supplementary Information

For detailed Supplementary Information, please refer to the associated PDF document.

## Tables

**Table 1 Model Performance in Temperature Prediction.**

| Model | Parameters(M) | MSE | MAE | $R^2$ |
|---|---|---|---|---|
| **ARIMA** | $7\times10^{-6}$ | **0.010** | **0.084** | **0.928** |
| **SARIMA** | $7\times10^{-6}$ | **0.010** | **0.084** | **0.928** |
| **LSTM** | **0.012** | **0.002** | **0.041** | **0.963** |
| **CNN-LSTM** | **0.017** | $3\times10^{-6}$ | **0.002** | **0.9999** |